\documentclass[sn-nature]{sn-jnl}

\usepackage{graphicx}%
\usepackage{multirow}%
\usepackage{amsmath,amssymb,amsfonts}%
\usepackage{amsthm}%
\usepackage{mathrsfs}%
\usepackage[title]{appendix}%
\usepackage{xcolor}%
\usepackage{textcomp}%
\usepackage{manyfoot}%
\usepackage{booktabs}%
\usepackage{algorithm}%
\usepackage{algorithmicx}%
\usepackage{algpseudocode}%
\usepackage{listings}%
\usepackage{xspace}
\usepackage{mathabx} 

\theoremstyle{thmstyleone}%
\theoremstyle{thmstyletwo}%

\theoremstyle{thmstylethree}%
\newtheorem{definition}{Definition}%

\newcommand{\NOME}{\textit{text}$_2$\textit{SQL}$_4$\textit{PM}\xspace}

\begin{document}

\title[Article Title]{Text-to-SQL Oriented to the Process Mining Domain: A PT-EN Dataset for Query Translation}


\author*[]{\fnm{Bruno Y.} \sur{Yamate}}\email{brunoyui@usp.br}
\equalcont{These authors contributed equally to this work.} 

\author[]{\fnm{Thais R.} \sur{Neubauer}}\email{thais.neubauer@usp.br}
\equalcont{These authors contributed equally to this work.} 

\author[]{\fnm{Marcelo} \sur{Fantinato}}\email{m.fantinato@usp.br}
\equalcont{These authors contributed equally to this work.} 

\author*[]{\fnm{Sarajane M.} \sur{Peres}}\email{sarajane@usp.br}
\equalcont{These authors contributed equally to this work.} 

\affil*[]{\orgdiv{Escola de Artes, Ciências e Humanidades}, \orgname{Universidade de São Paulo}, \orgaddress{\street{Av. Arlindo Béttio, 1000}, \city{São Paulo}, \postcode{03828-000}, \state{São Paulo}, \country{Brazil}}}




\abstract{This paper introduces \NOME, a bilingual (Portuguese-English) benchmark dataset designed for the text-to-SQL task in the process mining domain. Text-to-SQL conversion facilitates natural language querying of databases, increasing accessibility for users without SQL expertise and productivity for those that are experts. The \NOME dataset is customized to address the unique challenges of process mining, including specialized vocabularies and single-table relational structures derived from event logs. The dataset comprises 1,655 natural language utterances, including human-generated paraphrases, 205 SQL statements, and ten qualifiers. Methods include manual curation by experts, professional translations, and a detailed annotation process to enable nuanced analyses of task complexity. Additionally, a baseline study using GPT-3.5 Turbo demonstrates the feasibility and utility of the dataset for text-to-SQL applications. The results show that \NOME supports evaluation of text-to-SQL implementations, offering broader applicability for semantic parsing and other natural language processing tasks.} 



\maketitle

\section{Introduction}
\label{sec:intro}

The text-to-SQL conversion task \cite{yu_2018a_type_sql}, a specialized area of semantic parsing, involves generating SQL (Structured Query Language) statements from natural language utterances. By enabling access to database information through natural language holds the potential to democratize information retrieval, allowing users without knowledge of SQL commands or syntax to make basic queries \cite{Visperas2023}. Furthermore, solving the text-to-SQL task through semantic parsing enhances developer productivity by generating SQL statements that closely align with ideal ones. Recent advances in implementing such a task have leveraged modern deep neural networks \cite{Katsogiannis2023}, either complementing or replacing traditional rule-based parsing and mapping techniques. Even more recently, large language models (LLMs), coupled with diverse prompt engineering strategies, have demonstrated remarkable performance on benchmark datasets for the task \cite{yu_2018_spider_dataset}. Benchmark datasets used to evaluate solutions in this area are typically cross-domain and comprise multi-relational databases \cite{yu_2018_spider_dataset, zhong_2017_seq2SQL, li_2023_bird}. These datasets serve a dual purpose: they are employed both for model fine-tuning and for assessing the effectiveness of text-to-SQL strategies.

Although a range of domains is addressed in these benchmark datasets, they typically represent classical information retrieval domains such as cars, flights, music, sports, academia, and pets. While this diversity of domains provides valuable vocabulary and insights into information retrieval requirements for model training and evaluation, it is not comprehensive enough to develop models that perform effectively in more specialized domains. These domains often feature data structures with characteristics uncommon in relational databases and possess specific needs for information extraction. In this paper, we introduce a benchmark dataset designed to support information retrieval tasks relevant to the process mining domain. The study of applying the text-to-SQL task in process mining is motivated by the strong interdisciplinarity inherent to this field. Data analysts, process analysts, and organizational managers all play crucial roles in the practical implementation of process mining. By considering these professionals, we highlight the benefits of text-to-SQL solutions in two key areas: enhancing productivity in information retrieval for data analysts and improving accessibility to information for process analysts and organizational managers.

In the process mining domain, the data of interest is stored in event logs, where each record corresponds to process executions \cite{aalst_2016_process_mining, aalst_2022_process_mining}. These logs serve as a foundation for extracting valuable insights, whether through specialized algorithms, process query languages, or more commonly used computational strategies such as SQL and descriptive statistics. In this context, while event logs are not inherently required to be stored in a relational database, they can be transformed for SQL-based information retrieval. The most common standard for storing event logs in process mining is the XES (eXtensible Event Stream) format \cite{IEEE2016}, which, when converted for use in a relational database, generates a single, non-normalized table. This lack of normalization, combined with the specialized vocabulary and unique information needs of process mining, creates a domain in which text-to-SQL strategies tend to underperform when compared to classical domains. Despite the apparent simplicity of a context where information retrieval is performed via SQL from a single table in a relational database, exploratory studies have shown that querying information in this scenario can be quite challenging.

The benchmark dataset \NOME is introduced in this paper. \NOME is an open, bilingual (Portuguese and English), and annotated benchmark dataset, designed for training and evaluating text-to-SQL solutions within the specific context of process mining. To the best of our knowledge, no existing dataset includes natural language utterances, SQL statements, and annotations tailored to the process mining domain.
Thus, we assert that the contributions of this paper are:

\begin{enumerate}
    \item A complete description of the proposed bilingual (Portuguese-English) benchmark dataset with 1,655 natural language utterances, 205 corresponding SQL statements, and ten qualifiers\footnote{In this paper, a qualifier is defined as a label that represents a specific perspective of analysis in evaluating the competence of the text-to-SQL solution.}. The dataset includes 205 distinct utterances and 1,450 paraphrases, all formulated by humans (without the use of language models in the construction of the utterances or SQL statements). The qualifiers are employed to facilitate a contextualized analysis of the complexity embedded in the dataset. 
    \item An example of applying a text-to-SQL solution based on prompt engineering and the large language model GPT-3.5 Turbo, accompanied by a detailed and contextual analysis of the results with respect to the dataset's qualifiers. In this analysis, we highlight both the complexity of the task and the feasibility and utility of applying text-to-SQL for information retrieval in the process mining domain.
\end{enumerate}

The paper is structured as follows: Section \ref{sec:backRW} presents the fundamental theoretical concepts and related benchmark datasets; Section \ref{sec:dataset} describes the \NOME benchmark dataset, including the methods used in its construction, the qualifiers used for annotation, and statistics and examples of the instances included; Section \ref{sec:baseline} discusses the study of the text-to-SQL task supported by the dataset, establishing a baseline solution associated with it; finally, Section \ref{sec:conclusions} presents the conclusions, followed by the references.

\section{Background and Related Work}
\label{sec:backRW}

In this section, we present a brief definition of SQL and the text-to-SQL task, and some concepts of the process mining field, which is the domain of the dataset created. We also present related works that provide publicly available datasets similar to ours, whether for text-to-SQL tasks or process mining tasks.

\subsection{SQL and Text-to-SQL}
\label{sec:SQL}

Structured Query Language (SQL) is a declarative language used to organize, manipulate, and retrieve information from relational databases. This language is commonly employed to query data through a single request, known as an SQL statement, which consists of commands\footnote{The most common SQL commands are: SELECT with or without arithmetic operations (AVG, COUNT, MIN, MAX, SUM), INNER JOIN, WHERE, ALL, GROUP BY, HAVING, ORDER BY, BETWEEN, LIKE, IS NULL, IS NOT NULL, UNION, INTERSECT, EXCEPT, IN, NOT IN, ANY, SOME.} that allow customization of the retrieved data. An SQL statement operates on the relations defined in a schema of a relational database and returns, as a result, a relation (temporary, existing at runtime) in relational format. 

\vspace{1em}
\begin{definition}[\textbf{Schema, Relation, Relational Database \cite{elmasri2010}}]~
    \normalfont
    A relational database $\mathcal{D}$ is a collection of data organized according to the principles established in the relational data model, i.e., data organized as a collection of relations defined according to a schema. A relation schema $R(A_1, A_2, ..., A_n)$, consists of a relation name $R$ and a list of attributes, $A_1, A_2, ..., A_n$. Each attribute $A_i$ has a domain $dom(A_i,R)$ in the relation schema $R$. A relation $r$ of the relation schema $R(A_1, A_2, ..., A_n)$ is a set of $n$-tuples $r = \{t_1, t_2, ..., t_m\}$. Each $n$-tuple $t$ is a list of $n$ values $t = \langle v_1, v_2, ..., v_n \rangle$, ordered as defined by the list of attributes $A_1, A_2, ..., A_n$ of the relation schema and according with the corresponding $dom(A_i,R)$.
\end{definition}
\vspace{1em}

Since SQL is a language for computational processing, it follows a well-known syntax based on a standard norm.

\vspace{1em}
\begin{definition}[\textbf{The syntax of a standard SQL statement}]~
    \normalfont
\begin{itemize}
\small
    \item [] \textit{SELECT} $A_1$, $A_2$, ..., $A_n$ 
    \item [] \textit{FROM} source relations $r$
    \item [] \textit{WHERE} condition(s)
    \item [] \textit{GROUP BY} $A_1$, $A_2$, ..., $A_n$ 
    \item [] \textit{HAVING} condition(s)
    \item [] \textit{ORDER BY} $A_1$, $A_2$, ..., $A_n$  
\end{itemize}
\end{definition}

Each clause of an SQL statement has a well-established function: \textit{SELECT} is responsible for defining the attributes that should structure the resulting relation; \textit{FROM} indicates the relations that serve as data sources for information retrieval; \textit{WHERE} is a clause that applies filters to the data sources; \textit{GROUP BY} is responsible for grouping data, usually to support statistical summaries; \textit{HAVING} is a clause that filters the information from aggregations; and \textit{ORDER BY} is a clause used to impose sorting on the data in the resulting relation. In addition to this basic structure, the following are also cited as basic commands: arithmetic operations, set operations, string processing operations, and nullity test conditions.

The text-to-SQL task aims to generate a SQL statement from a natural language utterance that can be executed on a database to retrieve information. 

\vspace{1em}
\begin{definition}[\textbf{Text-to-SQL task}]~
    \normalfont
    Let $c: U \rightarrow S$ be the implementation of text-to-SQL task, in which U is a universe of natural language utterances, S a universe of SQL statements, and $c(u) \rightarrow s $ is a conversion procedure from elements $u \in U$ to elements $s \in S$. In this conversion procedure, the following assertions are valid, due to the expressiveness of the natural language and the SQL language:
    \begin{itemize}
    \small
     \setlength{\itemindent}{0.5cm}
        \item $U' \subset U \, | \, \forall \, (u'_{i}, u'_{j}) \in U'$ and $i \neq j, \, u'_{i}$ and $u'_{j}$ are paraphrases in natural language, meaning they express the same intent;
        \item $S' \subset S \, | \, \forall (s'_{i}, s'_{j}) \in S'$ and $i \neq j, \, s'_{i}$ and $s'_{j}$ are equivalent SQL statements, meaning that they produce the same result when executed;
        \item if $c(u'_{i}) \to s'_{i}$, then $c(u'_{i}) \to s'_{j}$, $c(u'_{j}) \to s'_{i}$ and $c(u'_{j}) \to s'_{j}$ for any $i,j$.
    \end{itemize}
\end{definition}

Solutions for implementing $c: U \rightarrow S$ aim to efficiently address two well-known problems: \textit{schema representation} for $R$, which serves as input for the conversion procedure $c$ along with the utterance $u$; and \textit{schema linking},  which involves connecting the terms and concepts in the utterance $u$ with the relations, attributes, and values in $D$. Building on the discussions outlined by Wang et al. \cite{wang_2020_ratsql}, we assert that the former problem entails an appropriate encoding of schema elements (such as relation names, attribute names, their respective domains, and primary and foreign keys), while the latter involves aligning entity references extracted from $u$ with the encoded schema elements.

Initially, the deep learning models fine-tuned for implementing $c$ treated schema representation and schema linking as separate problems, striving to devise sophisticated methods for representing database schemas, e.g. through graphs \cite{bogin_2019_schema_representation}. However, with the advent of the cross-domain text-to-SQL task, these two challenges began to be addressed \textit{in tandem} \cite{wang_2020_ratsql}\cite{li_2023_resdql}. The emergence of large language models for the text-to-SQL task has rendered this separation nearly imperceptible. When the utterance $u$ and database schema 
$R$ are provided via a prompt, implementations using these models aim to efficiently extract and present only the schema information from $R$ that is most relevant to the $u$ specified in the prompt \cite{dawei_2023_dail_sql}\cite{pourreza_2023_dinsql}.

\subsection{Process Mining Domain}
\label{sec:pm}

Process mining brings together data and process sciences with the main goal of automatically extracting knowledge about business processes from \textit{event logs}. An event log $L$ is a sequential file that records \textit{events} related to the execution of \textit{activities} (i.e., well-defined steps in a process) within the business process under analysis. Typically, we assume that each \textit{event} $e$ is related to a particular process instance, referred to as \textit{case} $c \in C$. Additional information, called \textit{attributes}, may be recorded in event logs, such as the timestamp of the event, the person or resource executing the activity, or any other data elements recorded with the event \cite{aalst_2016_process_mining}. 

\vspace{1em}
\begin{definition}[\textbf{Event, Attribute \cite{aalst_2016_process_mining}}] 
    \normalfont
    Let $\mathcal{E}$ be a universe of events, i.e. the set of all possible event identifiers. Events may have various attributes, such as timestamp, activity, resource, cost, and others. Let $\mathcal{AN}$ be a set of attribute names. For any event $e \in \mathcal{E}$ and name $n \in \mathcal{AN}: \#_n(e)$ is the value of attribute n for event $e$. Typically, for each existing attribute related to the events in $\mathcal{E}$, a domain is defined for its values, i.e., if we consider the \textit{timestamp} attribute $\#_{timestamp}(e)$, we define the time domain $T$ and $\#_{timestamp}(e) \in T, \forall e \in \mathcal{E}$.
\end{definition}
\vspace{1em}

Each event in the event log is globally unique, i.e., the same event cannot occur twice in a event log. An event log consists of \textit{cases} and cases consist of events. The events for a case are organized in a \textit{trace}, i.e., a sequence of unique events. Moreover, cases, like events, can have attributes. 

\vspace{1em}
\begin{definition}[\textbf{Case, Trace, Event log \cite{aalst_2016_process_mining}}]
    \normalfont
    Let $\mathcal{C}$ be the case universe, i.e. the set of all possible case identifiers. Cases, like events, have attributes. For any case $c \in \mathcal{C}$ and name $n \in \mathcal{AN}: \#_n(c)$ is the value of attribute $n$ for case $c$. Each case has a special mandatory attribute \textit{trace}: $\#_{trace}(c) \in \mathcal{E}^*$. $\hat{c} = \#_{trace}(c)$ is a shorthand for referring to the trace of a case.
    A \textit{trace} $\sigma \in \mathcal{E}^*$ is a finite non-empty sequence of unique events ascending ordered by occurrence time, i.e., for $1 \leq i < j \leq |\sigma|: \sigma(i) \neq \sigma(j)$ and $\sigma(i)$ occurs before $\sigma(j)$. An event log $L$ is a set of cases $L \subseteq \mathcal{C}^*$ where each event appears only once in the log, i.e. for any two different cases the intersection of their events is empty.
\end{definition}
\vspace{1em}

A definition that simplifies the concept of a trace is presented by Aalst \cite{aalst_2016_process_mining} as a `simple trace', and for practical purposes in process mining, `simple trace' is referred to as `variant' in the computational tools of the field. Both are used to represent an analytical perspective based on the control flow of a process instance, i.e., based on the notion of the execution of a sequence of activities. 

\vspace{1em}
\begin{definition}[\textbf{Simple trace, Variant}]
    \normalfont
     A \textit{simple trace} $\sigma'$ is a sequence of activities related to the events that make up the trace $\sigma$ \cite{aalst_2016_process_mining}. The term \textit{variant} is commonly used to refer to $\sigma'$ in computational tools. 
\end{definition}
\vspace{1em}

Table \ref{tab:log_example} shows an excerpt of an event log. Only three cases are shown, and their respective traces contain three, five, and four events, respectively. Each event has a unique identifier and several attributes. For example, the first event in the event log is an instance of the activity `Create ticket' that occurred on
February 20th at 10:30 was executed by Joana and cost 10 dollars. The second case starts with the third event in the event log and also refers to an instance of the activity `Create ticket'.

\begin{table}[htp]
\caption{Example of a log excerpt}\label{tab:log_example}
\centering
\begin{tabular}{ccllllr}
\hline
&&\multicolumn{5}{l}{Attributes} \\
\cmidrule{3-7}
 Event &Case \\
identifier& identifier &Activity &Timestamp &Resource &Cost &...\\
\midrule
1 &1 &Create ticket &20/02/2021 10:30 &Joana &10 &...\\
2 &1 &Activate ticket &20/02/2021 10:33 &Paul &50 &...\\
3 &2 &Create ticket &20/02/2021 10:33 &Ana &10 &...\\
4 &2 &Activate ticket &20/02/2021 10:40 &Paul &50 \\
5 &1 &Await for user input &20/02/2021 11:10 &Cris &100 &...\\
6 &2 &Await for user input &20/02/2021 15:50 &Cris &100 &...\\
7 &3 &Create ticket &23/02/2021 16:01 &Joana &10 &...\\
8 &3 &Activate ticket &23/02/2021 16:09 &Paul &50 &...\\
9 &3 &Handle ticket &23/02/2021 16:12 &Paul &50 &...\\
10 &3 &Close ticket &23/02/2021 16:55 &Paul &10 &...\\
11 &2 &Handle ticket &25/02/2021 11:42 &Cris &50 &...\\
12 &2 &Activate ticket &25/02/2021 12:40 &Cris &70 &...\\
... &... &... &... &... &... &...\\
\hline
\end{tabular}
\end{table}

Several process mining tasks and techniques rely on the sequence of the events within the cases (e.g. process model discovery, conformance checking, process monitoring, and descriptive, predictive, and prescriptive analysis). Therefore, the events within a case must be ordered on the event log by the moment of occurrence \cite{aalst_2016_process_mining}. In Table \ref{tab:log_example}, the `Timestamp' column provides the information of the moment when the activity was executed and the event log is ordered by it. It is also common to organize the event log grouping the events of each case, but the order of events within the case is always maintained. 

The field of process mining has certain particularities that should be noted here, as they can influence the declarative extraction of information procedures:

\begin{itemize}
    \item Partial ordering in event logs: In some information systems where event logs are recorded, no timestamp information is available. In other cases, timestamps can be too coarse, such as commonly seen in hospitals where information systems only record a date. In addition, when the event log results from merging data from different sources, timestamp-related problems can arise due to multiple clocks and delayed recording. However, in principle, such event log ordering does not require timestamps. One way to address ordering problems is to assume only a partial ordering of events (i.e., not a total order) and subsequently use dedicated process mining algorithms for this. Another way is to define the order based on domain knowledge or frequent patterns across days \cite{aalst_2016_process_mining}.
    \item Process query language: While the tabular nature of an event log allows the use of SQL, it is not the only declarative way to query information in an event log. Specialized languages for querying process data, called process query language (PQL) have been developed to offer process mining-specific operators.  For example, PQL includes the SOURCE and TARGET operators to link events registered in different tuples in a relation, the VARIANT operator to aggregate cases event names into a single string, and the CONFORMANCE operator for conformance checking. Despite these specialized languages, choosing SQL for use promotes generalization and accessibility, due to its popularity. SQL’s widespread use and familiarity far outweigh the benefits of a more specialized language like PQL, whose expertise is harder to find.
\end{itemize}

\subsection{Related work}

Several domain-specific datasets with pairs of English natural language utterances and the corresponding SQL statement exist and have been used by research community for decades, like ATIS \cite{dahl_1994_atis_dataset}, GeoQuery \cite{zelle_1996_geoquery}, Scholar \cite{iyer_2017_scholar}, Yelp and IMDB \cite{yaghmazadeh_2017_yelp_imdb}. 

With advancements in deep learning models and language models that generate a SQL statement given a natural language utterance, more robust datasets were published. Such datasets are considered cross-domain. In this context, `cross-domain' refers to a dataset that consists of multiple databases, each with schemas and data related to distinct domains. For the text-to-SQL task, the training and testing sets should not share the same domain; in other words, models are expected to generalize across different domains. Examples of datasets in such category include: 

\begin{itemize}
    \item WikiSQL \cite{zhong_2017_seq2SQL}: a large dataset consisting of 80,654 pairs of utterances and corresponding SQL statements on 24,241 databases. It is organized with only one relation and the SQL statements structure is very simple with just SELECT and WHERE clauses. 
    \item Spider \cite{yu_2018_spider_dataset}: a robust dataset consisting of 10,181 utterances and 5,693 corresponding SQL statements on 200 databases covering 138 different domains. This database contains multiple relations and more complex SQL statements, including nested clauses and operators such as JOIN, GROUP BY, ORDER BY, and HAVING. It also incorporates more 1,659 utterances in the training dataset that come from other datasets.
    \item BIRD \cite{li_2023_bird}: a robust dataset containing 12,751 utterances and SQL statements pairs across 95 databases with seven tables on average on 37 different domains. The SQL statements contains function operators such as DATE, YEAR, IIF, STRFTIME and CAST and the use of CASE conditions on SELECT. The utterances are challenging requiring knowledge domain to be answered. This dataset focuses on database values, with the databases containing dozens of rows, unlike other datasets that contain only a few rows.
\end{itemize}

When it comes to datasets in Portuguese for the text-to-SQL task, mRAT-SQL+GAP \cite{archanjo_2021_mRATSQL_GAP} is the only general-purpose dataset found. It is the result of a translation using Google Cloud Translation API\footnote{Cloud Translation API: \url{https://googleapis.dev/python/translation/latest/index.html}.} of 8,659 training utterances and 1,034 dev utterances (used for test purposes) from the Spider dataset. The SQL statements were kept in English. In the realm of process mining, Barbieri et al. (2022) presented a dataset with statements in the form of questions, in a style similar to the datasets used in question-answering tasks. Each question addresses a common information need in process mining, which can be resolved either through simple SQL statements or through complex specialized algorithms \cite{barbieri_2022_conversation_pm}. The dataset was used in the task of translating a natural language question to a logical query that could be run on existing process mining tools and it does not have gold standard outputs to be used in supervising learning methods. The original set of 250 questions was written in Portuguese and volunteers translated them to English, resulting in a total of 794 questions. Only the English questions are publicly available\footnote{\url{https://ic.unicamp.br/~luciana.barbieri/pmquestions.csv}}.

The dataset \NOME, introduced herein, differs from previous ones by featuring a unique set of characteristics: it is domain-specific (process mining), bilingual, designed for supervised learning methods, fully generated and reviewed by humans, and comes with a set of qualifiers and a baseline text-to-SQL solution based on a large language model. Although it is still small compared to datasets used for benchmarking large language models, it includes statements and SQL queries that cover a complete range of basic SQL commands, varying from simple statements to highly complex ones.

\section{\NOME Dataset}
\label{sec:dataset}

The dataset introduced in this paper, entitled \NOME, was designed to facilitate the development of natural language to SQL conversion engines, particularly those based on machine learning, for the field of process mining. Additionally, it serves as a benchmark for assessing any conversion engine designed to tackle the Text-to-SQL task within the process mining domain.

The contents of this dataset\footnote{The dataset, as well as excerpts from the event log, containing content in Portuguese and English, enabling the exploration of the dataset's SQL statements, are available at \url{https://github.com/pm-usp/text-2-sql}. The excerpts from the event log have been slightly modified from the original event log to facilitate a more accurate analysis of the correctness of the SQL statements.} include utterances for information extraction from a business process context, written in Portuguese and English, SQL statements that solve the requested information retrieval, and descriptive information that allows for generating some statistics about the dataset. The dataset is located in the domain of a business process concerning the dynamics of authorization requests for reimbursement of academic travel expenses incurred by university staff. The event log associated with this business process is well-known in the process mining community and was the subject of the BPI Challenge competition in 2020\footnote{\url{https://icpmconference.org/2020/bpi-challenge/}}.

This section presents the method followed to construct the \NOME dataset and a series of descriptive statistics that describe its complexity.

\subsection{Method}
\label{sec:methodD}

\begin{figure}[htp]
  \centering
  \includegraphics[width=\textwidth]{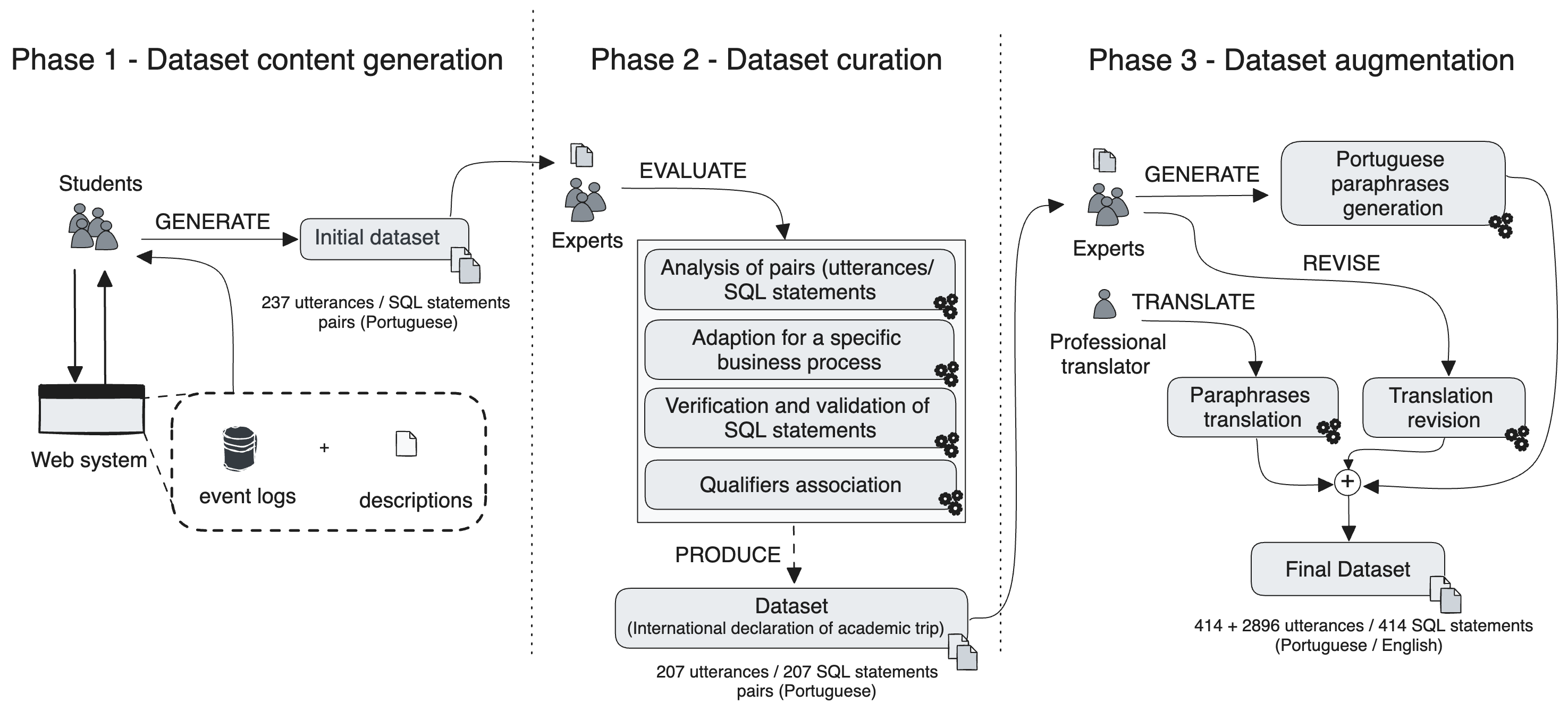}
  \caption{Overview of \NOME dataset generation process}
  \label{fig:dataset_overview}
\end{figure}

The three-phased method followed in the generation of the dataset is depicted in Figure \ref{fig:dataset_overview}. In Phase 1, initial dataset content was generated by undergraduate and graduate students as part of their coursework in Data Mining and Process Mining classes. A scoring system was implemented to incentivize students to produce high-quality content for this exercise. Subsequently, Phase 2 and Phase 3 were conducted by three researchers with extensive expertise in process mining and SQL. These phases involved an evaluation of the initial content, a domain adaptation of the complete dataset, and a data augmentation process. The remaining of this section provides a detailed description of each phase.
 
\vspace{0.2cm}
\noindent{\textbf{\textit{Phase 1 - Dataset content generation:}}} 

The goal of the first phase was the generation of the initial content, maximizing the variability of utterances and SQL statement types that would compose the \NOME dataset. To achieve this, the following strategies were applied:

\begin{itemize}
    \item \textbf{Participants}: 29 undergraduate students and 13 graduate students, enrolled in courses whose syllabus focused on process mining topics and who had previous knowledge of SQL, were invited to participate in an exercise to generate pairs (NL-PT utterance/statement SQL)\footnote{NL-PT: natural language in Portuguese.} that could appropriately express the extraction of useful information about a business process from an event log.
    \item \textbf{Score system and guidelines}: A scoring scheme was assigned to the exercise, considering criteria related to the quantity of pairs created, their correctness and the variety of commands used in the SQL statements (see Section \ref{sec:SQL}). Two guidelines were provided to the students to enhance the quality of their content: i) the utterances must be useful for a process manager to request information from the event log, and ii) only pure SQL statements should be elaborated, meaning that cursors and control flow could not be used. 
    \item \textbf{Business process}: Six different business processes, each represented by a descriptive text and at least one excerpt from an associated event log, were randomly presented to the students.\footnote{The business processes and their respective event logs were related to: incident management \cite{amaral_2019_enhancing_time_prediction}, open and close problems management \cite{BPI2013}, judicial performance of Brazilian justice \cite{Unger2021}, financial loan requests \cite{BPI2017}, authorization requests for civil construction \cite{BPI2015}, and international declaration of academic trip \cite{dongen_2020_bpi_challenge}.} Thus, different real-world situations were available for the creative process of the student group, increasing the chances of generating statements with different information extraction objectives.
    \item \textbf{Data collection}: An \textit{ad hoc} WEB system guided the students in producing the pairs  (NL-PT utterance/statement SQL). Through this system, students had access to three (of six) descriptions of business processes and one of their respective excerpts from event logs. From this information, they were required to register at least one utterance-statement pair for each business process. Additionally, non-normalized database tables\footnote{
    The event log schema consists of a single non-normalized table, lacking foreign keys, and containing duplicated information in the `id-case' column.}, each containing an excerpt from an event log, were made available within a relational database management system for students to test the SQL statements produced. A total of 237 utterance-statement pairs were produced through this exercise for the initial dataset content.
\end{itemize}

\vspace{0.2cm}
\noindent{\textbf{\textit{Phase 2 - Dataset curation:}}} 

The goal of the second phase was the improvement of the dataset content. The initial content of the dataset underwent detailed data curation involving three process mining experts (a senior researcher, a doctoral student, and a master's student with experience in the software industry hereafter referred to as \textit{expert A}, \textit{expert B}, and \textit{expert C}, respectively) in the following actions:

\begin{itemize}
    \item Analysis of each NL-PT utterance aimed at evaluating if they were correctly formulated and meaningful in the context of information retrieval in process mining, conducted by \textit{expert A} and \textit{expert C}. Correctness was evaluated regarding the appropriate use of process mining jargon concerning its basic concepts (see Section \ref{sec:pm}). Meaningfulness was evaluated concerning the usefulness of the information to be retrieved from that utterance. If an error or inadequacy was observed, either the utterance and its respective SQL statement were adapted, or the pair under analysis was discarded.
    \item The adaptation of the entire dataset to focus on a specific business process, carried out by \textit{expert A} and \textit{expert C}. Although several business processes were used in the content generation of the dataset (phase 1), we determined that only one of them would be the focus of attention in the dataset to enable its use in exploring text-to-SQL issues in the process mining domain, isolating problems that could arise from the variability of vocabulary resulting from the use of different business processes. The utterances and SQL statements were adapted according to the vocabulary used in the business processes related to authorization requests for reimbursement of academic travel expenses \cite{dongen_2020_bpi_challenge} (see examples in Table \ref{tab:adaption_domain}). 
    \item Verification and validation of the SQL statements by \textit{expert A} and \textit{expert C}, aiming to ensure correctness in the use of the SQL language and accuracy in information retrieval, according to what was specified in the utterance.
\end{itemize}

\begin{table}[htp]
  \caption{Examples of adaptation applied to the dataset in order to align it with the domain of international academic travel declarations. The changes are shown in bold. Note that this concerns the alteration of vocabulary inherent to the business process.}
  \label{tab:adaption_domain}
  \begin{tabular}{cp{5cm}p{5cm}}
    \toprule
    &Original & Adapted\\
    \midrule
    Example 1&\textit{Utterance}: How many activities \textbf{`closed'} do we have?&\textit{Utterance}: How many activities \textbf{`end trip'} do we have?\\
    &\textit{SQL}: SELECT count(*) FROM events\_log WHERE activity = \textbf{`Closed'}&
    \textit{SQL}: SELECT count(*) FROM events\_log WHERE activity = 
    \textbf{`End trip'}\\
    &&\\
    Event log: & incident management & international declaration of academic trip \\
    \midrule
    Example 2&\textit{Utterance}: How many events are associated with the \textbf{`Petition joined'} activity?&\textit{Utterance}: How many events are associated with the \textbf{`declaration rejected by director'} activity?\\
    &\textit{SQL}: SELECT COUNT (*) FROM events\_log WHERE activity = \textbf{`Petition joined'}&
    \textit{SQL}: SELECT COUNT (*) FROM events\_log WHERE activity = \textbf{`Declaration rejected by director'}\\
     &&\\
     Event log: & judicial performance of Brazilian justice & international declaration of academic trip \\
    \bottomrule
\end{tabular}
\end{table}

The dataset curation also involved associating the utterance-statement pairs with a series of qualifiers, carried out by the \textit{expert} B and \textit{expert C}, in order to organize the dataset's content regarding relevant aspects from the perspective of process mining (PMp), natural language (NLp), and Structured Query Language (SQLp):

\begin{itemize}
    \item \textbf{Qualifier 1} (PMp): refers to whether the utterance can be answered considering each event independently or if the events need to be aggregated by case. Values: \textit{event level - case level}.
    
    \item \textbf{Qualifier 2} (PMp): it refers to the perspective from which the process is analyzed given the information request in the utterance, and whether it is a statistical information extraction or conformance verification. Values: \textit{perspective (control flow, temporal, resource, cost) - descriptive statistics (control flow, temporal, resource, cost) - conformance}.
      
    \item \textbf{Qualifier 3} (PMp): it refers to the process mining concepts that occur as the objective of externalizing information, i.e., concepts that are used in the SELECT clause. If an aggregation occurs, then it is also indicated along with the aggregate concept. Values: \textit{case - event - timestamp - activity - resource - cost}.
 
    \item \textbf{Qualifier 4} (PMp): it refers to the process mining concepts that occur as the objective of filtering information, i.e., concepts that are used in the WHERE clause. Values: \textit{case - timestamp - activity - resource - cost - none}.
  
    \item \textbf{Qualifier 5} (NLp): it refers to the classic wh-question classification of the utterances. Values: \textit{how - what - which - when - who - none}.

   \item \textbf{Qualifier 6} (SQLp): it determines whether the SQL statement involves an aggregation function in the SELECT clause. Values: \textit{aggregation - none}.

    \item \textbf{Qualifier 7} (SQLp): it determines whether a condition on the GROUP BY clause is required to answer the utterance, i.e., it indicates the presence of a HAVING clause. Values: \textit{having - none}.
    
    \item \textbf{Qualifier 8} (SQLp): it refers to the hardness criteria provided by Spider\cite{yu_2018_spider_dataset}. The hardness criteria consider four levels of difficulty based on number of SQL components (selections and conditions) present in SQL statements. Values: \textit{easy - medium - hard - extra hard - no hardness}.

\end{itemize}

After curatorial actions were performed, the dataset consisted of 205 revised and qualified utterance-statement pairs in Portuguese.

\vspace{0.2cm}
\noindent{\textbf{\textit{Phase 3 - Dataset augmentation:}}} 

The Phase 3 of dataset generation was dedicated to data augmentation. Two actions were carried out in this phase: building paraphrases and translating the utterances/statements into English. The paraphrases were created manually by the \textit{expert A} and \textit{expert B}. For paraphrase creation, the original utterances were either completely rewritten or had some elements replaced by linguistic or technical synonyms. Table \ref{tab:strategies_paraphrase_creation} shows three cases of paraphrase creation, illustrating the three strategies used.

\begin{table}[htp]
  \caption{Strategies used for paraphrase creation}
  \label{tab:strategies_paraphrase_creation}
  \begin{tabular}{p{4cm}p{4cm}l}
    \toprule
    Base utterance&Paraphrase&Strategy\\
    \midrule
    Which cases had their first log record before March 2017?&Processing records began occurring before March 2017 for which declarations?&completely rewriten\\
    &&\\
    Which cases \textbf{arrived at} the `end trip' activity between 2016 and 2017?&Which cases \textbf{went through} the `end trip' activity between 2016 and 2017?&replace by linguistic synonyms\\
    &&\\
    Which \textbf{cases} went through the `declaration rejected by supervisor' activity? Sort ascending by start date.&Which \textbf{process instances} went through the `declaration rejected by supervisor' activity? Sort the answer in ascending order by start date.&replace by technical synonyms\\
    \bottomrule
\end{tabular}
\end{table}

Once the paraphrases were created, two additional qualifiers were established, one regarding the process mining perspective and the other regarding the natural language perspective:

\begin{itemize}
    \item \textbf{Qualifier 9} (PMp): The meaning of each value for this qualifier is as follows. Values: \textit{value- generic - domain}. 
        \begin{itemize}
            \item \textit{value}: the vocabulary of the utterance has an explicit connection with the schema of database tables/columns and column values, and in most cases, values are enclosed in single quotes in the utterance, allowing them to be easily replaced with other values.
            \item \textit{generic}: the vocabulary of the utterance has an explicit connection with the schema of database tables/columns and with values with specific columns (case ID, event ID and timestamp), therefore it is process domain independent.
            \item \textit{domain}: the vocabulary of the utterance primarily considers the natural vocabulary used in the process domain and has little or no explicit connection with the schema of database tables/columns and column values.
    \end{itemize}
    \item \textbf{Qualifier 10} (NLp): it refers to derived utterances known as paraphrases from an initial utterance known as the base. Values: \textit{base - paraphrase}.
\end{itemize}

Since all individuals involved in generating the dataset are Portuguese native speakers, an English native professional translator was hired to translate all utterances into English. The translator was informed about the dataset's objective and was guided on the use of process mining field jargon and the importance of maintaining the integrity of the paraphrases during the translation. The translation was reviewed by \textit{expert C} to ensure that the process mining jargon and the original meaning of the utterances were maintained. 

To complete the process of creating the dataset in two languages, we established two event logs: one with the original values in English and another with the values translated into Portuguese. Consequently, the SQL statements also have two versions to align with their respective event logs. The decision to create both versions was important so that the evaluation of text-to-SQL conversion capability could be performed equally for both languages.

As a result, the dataset comprises 1,655 quadruples (NL-PT utterance, NL-EN utterance, SQL statement - PT, SQL statement - EN), consisting of 205 originals and 1,450 paraphrases associated with ten qualifiers. 

\subsection{Statistics and examples}
\label{sec:stat}

In this section we present statistics of the dataset according to the series of qualifiers defined in Section \ref{sec:methodD}. For each dataset analysis perspective (PMp, NLp, and SQLp), we present examples of utterances or statements that illustrate the classifications established under each qualifier, overviews of how many quadruples are associated with each class, and relationships between the classes of different qualifiers, whenever relevant.


\vspace{0.2cm}
\noindent{\textbf{\textit{Process mining perspective (PMp):}}}

Table \ref{tab:examples_qualifier_pmp} shows examples for the qualifiers 1, 2, 3, 4 and 9. 
Observing the examples for each qualifier mentioned in the table clarifies the interpretation used when they were applied to the dataset:
\begin{itemize}
    \item for Qualifier 1, the examples clarify what we consider as an information request at the event level or at the case level. In the former example, all filters (in this case, filters related to the activity name, resource name, and timestamp) are applied within each row of the table, meaning a specific event is being analyzed. In the latter example, it is necessary to analyze two subsequent rows, including the row where the filter on the activity name occurs, the analysis goes beyond the event to involve a case context; to make this possible in a standard SQL statement, the table (event log) is recursively wrapped in an INNER JOIN command.
    \item for Qualifier 2, the first example concerns a request in which a user is interested in a feature about the workflow of process instances, more specifically when it involves passing through an activity, meaning some perspectives of the process dynamics is being explored; furthermore, a restriction regarding the person (resourve) involved in the work is specified. The second example illustrates a request in which a simple count is requested. The third example can be seen as an audit activity (a conformance verification task), as it requests cases with the highest numbers of activities executed and their duration.
    \item for Qualifier 3, the utterance in the first example requests information about `events' that occurred within a specific time period; to address this, the SQL statement must include this information in the SELECT clause. In the second example, the focus is on `cases', thus, information about the `case' concept is associated with the SELECT clause.
    \item for Qualifier 4, in the first example, the utterance refers to event information related to a specific activity, so a filter related to the `activity' concept should be applied in the WHERE clause. In the second example, the process mining concept of interest pertains to the `resource' concept.   
    \item for Qualifier 9, in the first example, the text-to-SQL converter should identify which attribute, in the table schema, the explicitly mentioned values can be found. The second example is an utterance that applies to any event log, regardless the underlying business process domain of the event log. Finally, in the third example, specific domain words are used and they must be indirectly related to the table schema, so the text-to-SQL converter should understand that `interventions' refers to `events' and `declaration' refers to `case'.
\end{itemize}

\begin{table}[htp]
    \caption{Examples for qualifier categories from a process mining perspective}
    \label{tab:examples_qualifier_pmp}
    \centering
    \begin{tabular}{p{1.5cm}p{7cm}p{3.5cm}}
        \toprule
        &Example & Qualifier value\\
        \midrule
        Qualifier 1 & What were the activities carried out by the resource named Thomas in the first semester of 2018? &  event level \\
        &What are the activities that preceded a `send reminder' occurrence? &  case level \\
        &&\\
        Qualifier 2 &Show the identifiers of the cases that went through the `declaration rejected by pre-approver' activity, except those in which the activity was performed by the Douglas `resource'. & perspective \newline(control-flow and resource) \\
        &How many resources worked on each case? & descriptive statistics \newline(resource)\\
        &What were the five cases with the most performances of activities and what was the duration of each of them? & conformance\\
        &&\\
        Qualifier 3 & What events occurred in the year 2017? & event \\
        & In which declarations were up to 20 activity occurrences performed & case\\
        &&\\
        Qualifier 4 & Which resources were responsible for the `declaration rejected by pre-approver' activity? & activity \\
        &List the cases in which the `Wayne' resource was allocated, ordering the response by case identifier. & resource\\
        &&\\
        Qualifier 9 & How many times were `start trip' and `end trip' activities performed? & value\\
        &Which cases were entirely handled by the same resource? & generic\\
        &How many interventions were carried out in the processing of each declaration? & domain\\         
        \bottomrule
    \end{tabular}
\end{table}

\begin{figure}[htp]
  \centering
  \includegraphics[width=\linewidth]{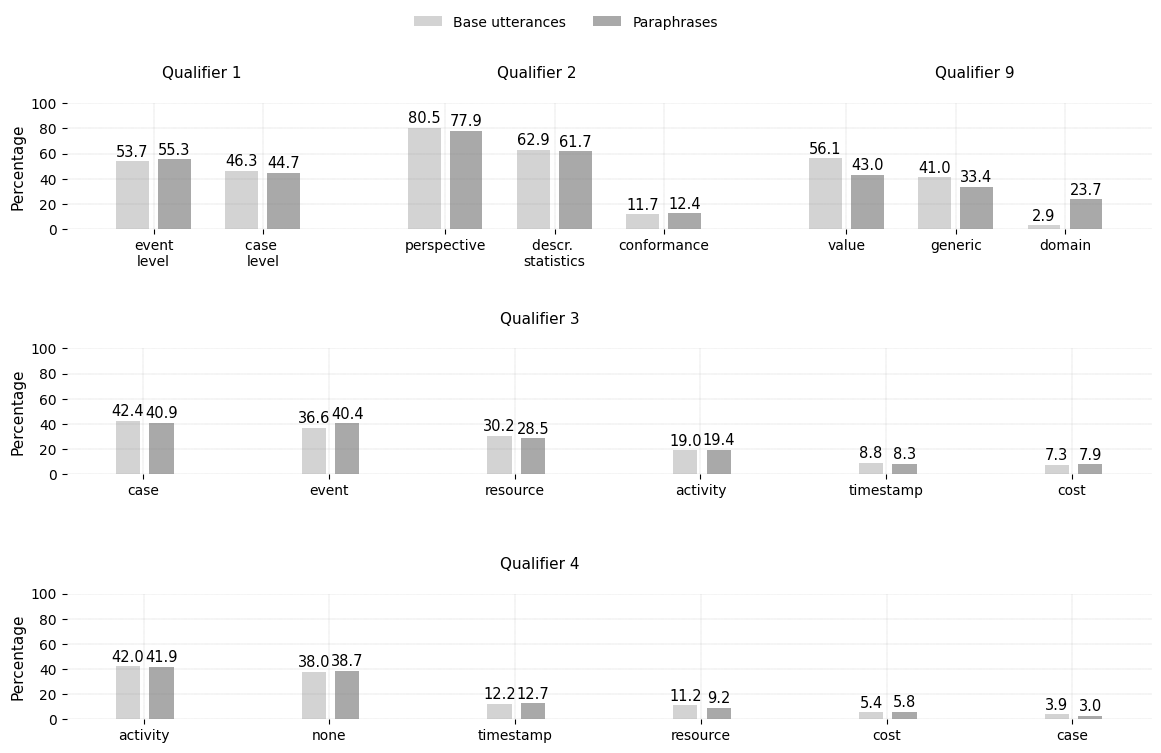}
  \caption{Process mining qualifiers comparison}
  \label{fig:process_mining_qualifiers}
\end{figure}

Table \ref{tab:pmp_table} shows the number of quadruples classified in each class of qualifiers 1, 2, 3, 4 and 9. The charts organized in Figure \ref{fig:process_mining_qualifiers} show the information about the distribution of the quadruples over the qualifiers. For qualifier 2, there is more than one value associated with the same quadruple, which causes the total number of associated values to be greater than the total number of quadruples in the dataset.

Some values for the qualifiers are infrequently present in the dataset. This is mainly due to the creative bias of the original utterances created by the students who participated in Phase 1 of the dataset generation process. Some highlights include the low number of requests that: are formulated with reference to the most likely discourse for business managers (value \textit{domain} for Qualifier 9); allude to compliance verification tasks; deal with cost analysis in general; deal with temporal analysis as information resulting from the SQL request; and deal with case-based filtering. 

Several factors may have contributed to the low occurrence. Among them, we noted: the students training involved in the utterance formulation being entirely focused on the computing area and not on business; the difficulty of considering auditing as information to be extracted from SQL requests; and the fact that some event logs used in the initial generation of utterances did not contain information about event cost. Specifically regarding the `domain' value in Qualifier 9, there is a higher number of utterances among the paraphrase subset. This is because \textit{expert A} created utterances with a vocabulary different from that commonly used by SQL programmers or process mining analysts. These utterances are composed of language that more closely resembles what would be used by business managers, considering domain-specific nomenclature rather than the standard terminology present in event logs and associated relational tables.

\begin{table}[htp]
  \caption{Process mining perspective: number of quadruples per qualifier/class. Legend: B - base; P - paraphrase; BP - base + paraphrase}
  \label{tab:pmp_table}
  \begin{tabular}{p{1.5cm}p{2.7cm}p{0.35cm}p{0.35cm}p{0.45cm}|p{1.35cm}p{1.2cm}p{0.3cm}p{0.3cm}p{0.4cm}}
    \hline
     & & B & P & BP & & & B & P & BP\\  
    \hline
    Qualifier 1  & event level & 110 & 802 & 912  & Qualifier 3 & case      & 87 & 593 & 680\\
                 & case level  & 95  & 648 & 743  & (SELECT)    & event     & 75 & 586 & 661\\
                 &             &     &     &      &             & resource  & 62 & 413 & 475\\
    Qualifier 2  & perspective & 165 & 1130 & 1295&             & activity  & 39 & 281 & 320\\
                 & descriptive statistics & 129 & 894 & 1023&   & timestamp & 18 & 120  & 138\\
                 & conformance & 24 & 180 & 204&                & cost      & 15  & 114  & 129 \\
    & & & & & & & & & \\                            
    Qualifier 9 & value   & 115 & 623  & 738 & Qualifier 4 & activity  & 86 & 608 & 694\\
    & generic & 84 & 484 & 568 & (WHERE) & none      & 78 & 561 & 639 \\ 
    & domain & 6 & 343 & 349 & & timestamp & 25 & 184 & 209\\
    & & & & && resource  & 23 & 133 & 156 \\
    &  &  &  &  & &cost      & 11 & 84  & 95 \\ 
    &  &  &  &  &&case      & 8  & 44  & 52\\ 
  
    \hline
\end{tabular}
\end{table}

\vspace{0.2cm}
\noindent{\textbf{\textit{Natural Language  (NLp):}}}

Table \ref{tab:examples_qualifier_nlp} shows some examples for the qualifiers 5 and 10. To apply the qualifications related to natural language processing, the following interpretations were assumed:

\begin{itemize}
    \item for Qualifier 5, the classification of utterances as wh-questions can result in three situations that warrant attention: an utterance may have an imperative form, as in the first example in Table \ref{tab:examples_qualifier_nlp}, in which case the value `none' is chosen; an utterance may involve more than one condition for requesting information, leading to the assignment of multiple classifications (second example in Table \ref{tab:examples_qualifier_nlp}); semantically equivalent utterances (including paraphrases) may have different linguistic constructions and therefore receive different classifications, as shown in the three last examples in Table \ref{tab:examples_qualifier_nlp}.
    \item for Qualifier 10, the examples demonstrate the construction of paraphrases based on conceptual equivalence (`to perform an activity' means `to be involved in in an event' and is related to `the employee's workload'). Other strategies include replacing date formats with written-out dates, changing interrogative sentences to imperative sentences, or using topicalization.
\end{itemize}

\begin{table}[htp]
    \caption{Examples for qualifier categories from the natural language perspective }
    \label{tab:examples_qualifier_nlp}
    \centering
    \begin{tabular}{p{1.5cm}p{7cm}p{3.5cm}}
        \toprule
        &Utterance & Qualifier value\\
        \midrule
        Qualifier 5&Show all the cases that ended in March 2018.& none\\
        &Which resources are related to more events and how many events are they related to? & which;how\\
        &&\\
        &Return the five resources requested in the greatest number of cases.& none\\
        &What are the top 5 resources with the most cases?& what\\
        &Report the five employees who made the most declarations.& who\\
         \midrule
        Qualifier 10 & How many times did the `Thomas' resource perform an activity in 2017? &  base \\
        &How many events was the `Thomas' resource involved in in 2017?& paraphrase\\
        &What was the workload of the employee Thomas in terms of performing actions for processing declarations in 2017?&paraphrase\\
        \bottomrule
    \end{tabular}
\end{table}

\begin{figure}[htp]
  \centering
  \includegraphics[width=\linewidth]{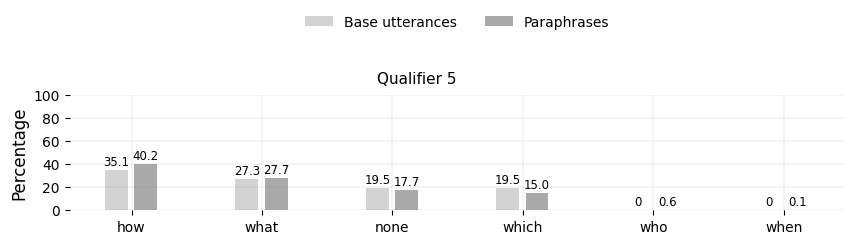}
  \caption{WH-question classification comparison}
  \label{fig:nl_type_qualifier}
\end{figure}

\begin{figure}[htp]
  \centering
  \includegraphics[width=\linewidth]{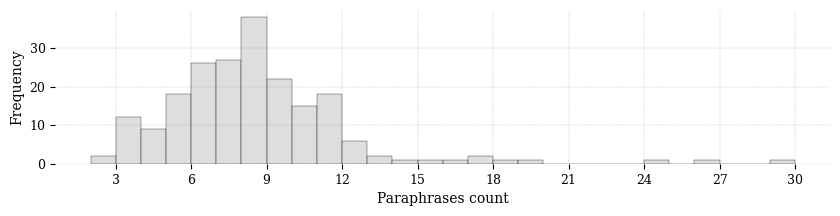}
  \caption{Paraphrases count distribution. X-axis: Number of paraphrase utterances per base utterance; Y-axis: Number of base utterances containing a specific number of paraphrase utterances. Example: there are 8 paraphrase utterances for 38 base utterances.}
  \label{fig:paraphrases_frequency}
\end{figure}

Table \ref{tab:nlp_table} shows the number of quadruples classified in each class of the Qualifier 5. The referenced distribution of paraphrases is shown in Figure \ref{fig:nl_type_qualifier}. The chart in Figure \ref{fig:paraphrases_frequency} shows the distribution of base and paraphrase utterances present in the dataset. In this distribution, we observe the number of paraphrases created for each base utterance, noting that most base utterances have between six to nine paraphrases. 

\begin{table}[htp]
  \caption{Natural language perspective: number of quadruples per values for Qualifier 5. The sum of the counts exceeds the number of utterances in the dataset because some receive more than one classification.}
  \label{tab:nlp_table}
  \begin{tabular}{p{1.4cm}p{1.4cm}ccc}
    \hline
     & & Base & Paraphrase & Base + Paraphrase \\
    \hline
     Qualifier 5 & how   & 72 & 401 & 473 \\
                 & what  & 56 & 257 & 313 \\
                 & none  & 40 & 583 & 623 \\
                 & which & 40 & 217 & 257 \\
                 & who   & 0  & 8   & 8   \\
                 & when  & 0  & 2   & 2   \\
    \hline
\end{tabular}
\end{table}

\vspace{0.2cm}
\noindent{\textbf{\textit{Structured Query Language perspective (SQLp):}}}
\vspace{0.2cm}

Table \ref{tab:examples_qualifier_sql} shows examples for the Qualifier 6, 7 and 8. The qualifier 6 indicates whether an aggregation operation is present or not in the SELECT clause, and Qualifier 7 specifies whether a filter is applied after the GROUP BY operation. The class assignment for the qualifier 8 was carried out through automatic analysis of SQL statements, according to the criteria adopted by the Spider dataset \cite{zhong_2020_semantic_evaluation}. Essentially, these criteria are based on the number of SQL components, selections, and conditions contained in the SQL statement. For example\footnote{For the complete set of rules, see the Spider \cite{yu_2018_spider_dataset}.},

\begin{itemize}
\item the easy class contains only one projection in the SELECT clause and no more than one condition in either the WHERE clause or GROUP BY HAVING clause, but not both; 
\item the medium class contains two projections in the SELECT clause and one condition on the WHERE clause with a GROUP BY clause; 
\item the hard class contains nested subqueries; 
\item the extra class contains GROUP BY HAVING clause with nested subqueries.   
\end{itemize}

\begin{table}[htp]
    \caption{Examples qualifier categories from a structured query language perspective}
    \label{tab:examples_qualifier_sql}
    \centering
    \begin{tabular}{p{1.5cm}p{8.5cm}p{2cm}}
        \toprule
        &Statement SQL & Qualifier value\\
        \midrule
        Qualifier 6 & SELECT \textbf{DISTINCT idcase} FROM event\_log WHERE cost $>$ 2225 & none\\
        & SELECT \textbf{count(*)} FROM event\_log WHERE resource = `Peter' & aggregate\\
        & & \\
        Qualifier 7 & SELECT COUNT(*), resource FROM event\_log \textbf{GROUP BY resource} & none\\
        & SELECT idcase, resource from event\_log group by idcase \textbf{HAVING COUNT(DISTINCT resource) = 1} & having\\
        & & \\
        Qualifier 8 & SELECT count(*) FROM events\_log WHERE activity = `End trip' & easy\\
        & SELECT resource, count(DISTINCT idcase) FROM events\_log WHERE timestamp BETWEEN `2017-01-01' AND `2017-05-31' GROUP BY resource & medium\\
        & SELECT COUNT (DISTINCT resource) FROM events\_log WHERE idcase IN (SELECT idcase FROM events\_log GROUP BY idcase HAVING COUNT(*) $>$ 3) & hard\\
        & SELECT idcase from events\_log GROUP BY idcase HAVING  COUNT(*) IN (SELECT COUNT(*) FROM events\_log  GROUP BY idcase ORDER BY COUNT(*)  desc LIMIT 10) ORDER BY COUNT(*) DESC & extra\\
        & WITH RankedEvents AS (SELECT id, activity, timestamp, LEAD(timestamp) OVER (PARTITION BY idcase ORDER BY timestamp) AS next\_timestamp, idcase FROM events\_log), DurationEvents AS (SELECT activity, (strftime(`\%s', next\_timestamp) - strftime(`\%s', timestamp)) as duration FROM RankedEvents WHERE next\_timestamp IS NOT NULL) SELECT activity, AVG(duration) AS average\_duration FROM DurationEvents GROUP BY activity& no hardness\\
        \bottomrule
    \end{tabular}
\end{table}

The `no hardness' class concerns the impossibility of classifying some gold SQL statements. This impossibility is due to limitations arising from the scope of SQL commands used in the Spider dataset. Such limitation refers to the nonexistence of the following SQL commands or operators: 

\begin{itemize}
\item `NULL' and `NOT NULL'; 
\item function \textit{strftime}; 
\item WITH clause; 
\item table alias on FROM clause used on SELECT (ex.: SELECT \textit{p.}attribute FROM table p ...); 
\item alias on FROM clause resulting from a SUBSELECT; 
\item clauses after a FROM clause with SUBSELECT; 
\item parentheses on WHERE clause; alias on SELECT clauses (ex.: SELECT attribute \textit{as c}); 
\item window functions (ex.: lead).
\end{itemize}

\begin{figure}[htp]
  \centering
  \includegraphics[width=\linewidth]{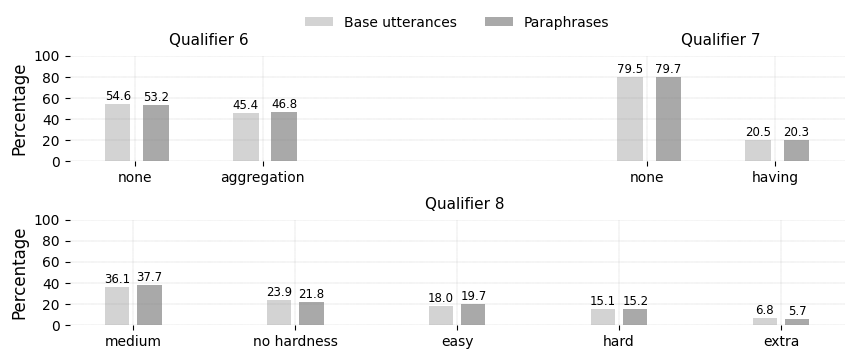}
  \caption{SQL qualifiers comparison}
  \label{fig:sql_qualifiers}
\end{figure}

The number of quadruples classified in each class of qualifiers 6, 7 and 8 is shown in Table \ref{tab:sql_table}. The respective distributions of these quadruples over the qualifier are illustrated by the charts organized in Figure \ref{fig:sql_qualifiers}.  

\begin{table}[htp]
  \caption{Structured Query Language perspective: number of quadruples per qualifier/class. Legend: B - base; P - paraphrase; BP - base + paraphrase}
  \label{tab:sql_table}
  \begin{tabular}{p{1.5cm}p{1.5cm}p{0.35cm}p{0.35cm}p{0.45cm}|p{1.35cm}p{1.8cm}p{0.3cm}p{0.3cm}p{0.4cm}}
    \hline
     & & B & P & BP & & & B & P & BP\\  
    \hline
    Qualifier 6  & none & 112  & 772 &  884 & Qualifier 8 & medium & 74 & 546 & 620 \\
                 & aggregation & 93 & 678 & 771 &             & no hardness & 49 & 316 & 365 \\
                 &  &  &  & &             & easy   & 37 & 285 & 322 \\
    Qualifier 7 & none & 163 & 1156 & 1319 &             & hard   & 31 & 220 & 251 \\
                 & having & 42 & 294 & 336 &             & extra  & 14 & 83 & 97 \\
   \hline
\end{tabular}
\end{table}

\section{\textit{text}$_2$\textit{SQL}$_4$\textit{PM} Baseline}
\label{sec:baseline}

This section presents the results of using the \NOME Dataset as a basic test case for implementing the text-to-SQL task. Based on these results, a baseline is established to assess the feasibility and potential value of applying this task in the process mining domain. The baseline is derived from the use of a large language model from the GPT family. The following sections provide a detailed overview of the model application method and the results obtained.

\subsection{Method}
\label{sec:methodB}

The following resources and procedures were applied in the test for building the \NOME baseline\footnote{All scripts and files used in the experiment are available in \url{https://github.com/pm-usp/text-2-sql}.}:

\vspace{0.2cm}
\noindent\textbf{\textit{Large Language Model}}: the GPT-3.5 Turbo model was used to provide SQL statements for each utterance in the dataset. The company OpenAI is one of the leaders in providing large language models and is the provider of the chosen model. The model used was selected for this purpose mainly due to its recognized performance in the text-to-SQL task \cite{dong_2023_c3}\cite{dawei_2023_dail_sql}  and its affordable cost.

\vspace{0.2cm}
\noindent\textbf{\textit{Database Management System (DBMS)}}: the DBMS SQLite we used for verification and validation of the SQL statements provided by the GPT-3.5 Turbo model. For this purpose, two database instances were generated for storing the event log `international declaration of academic trip' \cite{dongen_2020_bpi_challenge}, each containing a single relation (table) with attribute names in English and values in either Portuguese or English. In both cases the instance values are case insensitive. 

\vspace{0.2cm}
\noindent\textbf{\textit{Prompt Engineering}}: The first prompt used in OpenAI's official Text-to-SQL demo, known as the OpenAI Demonstration Prompt, was used. We tested prompts in both Portuguese and English. For Portuguese utterances, the entire prompt was translated to Portuguese, except the database schema, which was kept in English. Example of prompts in English are depicted in the Listing \ref{lst:prompt_ex_english}, and examples in Portuguese, in the Listing \ref{lst:prompt_ex_portuguese}. 
The prompts were designed using a zero-shot\footnote{According to Liu et at. \cite{Liu2023}, zero-shot prompt is a strategy where a pre-trained language model is applied to a task without any additional training specific to that task. The model uses predefined cloze or prefix prompts to generate the desired output, and this approach is called `zero-shot' because no task-specific training data is used.} approach, with: (i) the task specification, (ii) the database schema to be used in the task, and (iii) a text completion prompt that provides the first word of the model's response. 

    \definecolor{backcolour}{rgb}{0.96,0.96,0.96}

\lstset{ backgroundcolor=\color{backcolour}, texcl=true, basicstyle=\small\sf, breaklines = true}
\begin{lstlisting}[caption={Prompt example (English)}, label={lst:prompt_ex_english}]

    ### Complete sqlite SQL query only and with no explanation
    ### SQLite SQL tables, with their properties:
    #
    # event_log(id, activity, timestamp, resource, cost, idcase)
    #
    ### What events associated with the 'end trip' activity did not take place on December 12, 2017?
    SELECT
    
\end{lstlisting}

\lstset{ backgroundcolor=\color{backcolour}, texcl=true, basicstyle=\small\sf, breaklines = true}
\begin{lstlisting}[caption={Prompt example (Portuguese)}, label={lst:prompt_ex_portuguese}, literate={ç}{{\c{c}}}1{ã}{{\~a}}1{à}{{\`a}}1]

    ### Complete somente a consulta sqlite SQL e sem explicação
    ### Tabelas SQLite SQL, com suas propriedades:
    #
    # event_log(id, activity, timestamp, resource, cost, idcase)
    #
    ### Quais eventos associados à atividade `fim da viagem' não aconteceram no dia 12 de dezembro de 2017?
    SELECT 
\end{lstlisting}

\vspace{0.2cm}
\noindent\textbf{\textit{Evaluation}}: To evaluate the performance of the GPT-3.5 Turbo model on the text-to-SQL task using the \NOME dataset, two indicators were used: `exact set match without values' and `execution accuracy' \cite{zhong_2020_semantic_evaluation}, referred to here as the structure indicator and run indicator, respectively. 
These indicators are widely used in related literature.

    \begin{itemize}
        \item \textbf{Structure indicator}: 
        evaluates only the structure of the SQL statement, ignoring values in clauses condition. According to this indicator, a response is considered valid if the SQL statement generated by the model exactly matches a gold-standard SQL statement that was previously associated with the input utterance. This gold-standard statement serves as the reference for the correct information retrieval requested in the utterance. On the one hand, the success rate for this metric can be underestimated since a simple alias added on SELECT clause of the SQL statement generated by the GPT-3.5 Turbo model is considered a failure. Table \ref{tab:structure_indicator_example} shows examples on this matter. On the other hand, this metric does not validate the DISTINCT keyword on SELECT clause, which can result in an overestimation of success.
        In addition, due to limitations of the implementation used related to `no hardness' class, out of the 1,655 utterance-SQL statement pairs, 365 are not included in the calculation of the total result percentages for this metric, resulting in 1,290 pairs to be analyzed.

        \item \textbf{Run indicator}: evaluates the results of SQL statements when executed in a DBMS. An SQL statement generated by the model is considered valid if the results obtained by its execution match the results obtained by executing the gold SQL statement previously associated with the input utterance. For this indicator, a case-insensitive strategy was adopted for the values of attributes involved in the SQL statements conditions. 

    \end{itemize}

    \begin{table}[htp]
          \caption{Examples of failure on structure indicator just by simple addition of alias on SELECT clause of the SQL statement generated}
          \label{tab:structure_indicator_example}
          \begin{tabular}{cp{5cm}p{5cm}}
            \toprule
            &Generated & Gold\\
            \midrule
            English & SELECT resource, \textbf{COUNT(*) as num\_activities} FROM event\_log WHERE activity = `declaration approved by administration' GROUP BY resource & SELECT \textbf{count(*)}, resource FROM event\_log WHERE activity = `Declaration approved by administration' GROUP BY resource\\
            Portuguese & SELECT \textbf{COUNT(DISTINCT resource) AS total\_recursos} FROM event\_log & SELECT \textbf{COUNT(DISTINCT resource)} FROM event\_log \\
            \bottomrule
        \end{tabular}
    \end{table}
    
\subsection{Results}
\label{sec:app}

The results forming the proposed \NOME baseline are presented from four perspectives: performance under the structure indicator; performance under the run indicator; comparative performance analysis between the indicators; and general challenges.

\vspace{0.2cm}
\noindent\textbf{\textit{Structure indicator}}: Among the 1,290 utterance-SQL pairs of utterance-statements SQL analyzed for each language, the success rate for the structure indicator was 31.8\% for Portuguese and 32.7\% for English, corresponding to 410 and 422 correct answers, respectively. Some characteristic factor of the process mining domain influence the low success rates achieved by this indicator. Table \ref{tab:examples_utterance_complexities_cases_conformance} presents an example of such factors: 

\begin{itemize}
    \item the implicit reference to the event concept through its definition (activity execution), applied in the gold SQL statement, in contrast to the explicit reference to the attribute `activity' in the generated SQL statement. Although both SQL statements are equivalent, the indicator points to an error.
    \end{itemize}

\begin{table}[htp]
  \caption{Example of utterance complexity to which the structural indicator is sensitive. The differences between the gold and generated SQL statements are highlighted in blue.}
  \label{tab:examples_utterance_complexities_cases_conformance}
  \begin{tabular}{p{0.4cm}p{11.5cm}}
    \toprule
    EN & \textbf{Utterance}: Which process instances have more than 20 activities performed?\\
    PT & \textbf{Utterance}: \textit{Em quais instâncias de processo há mais de vinte ocorrências de atividades executadas?}\\
            & \textbf{Gold SQL}: \\
            &SELECT idcase FROM event\_log GROUP BY idcase HAVING count(*) $>$ 20\\
            & \textbf{Generated SQL - for both natural languages}: \\ 
            & SELECT idcase FROM event\_log GROUP BY idcase HAVING \textcolor{blue}{COUNT(DISTINCT activity)} $>$ 20 \\
    \bottomrule
\end{tabular}
\end{table}

However, the values the indicator yield different magnitudes depending on the group of utterance-SQL pairs being analyzed. Table \ref{tab:results_pmp_structure} presents detailed results for three PMp qualifiers. This closer, qualifier-level analysis enables us to identify where challenges arise within each category, offering insights not apparent from the overall success rates:

\begin{itemize}
    \item the lower success rate observed in the `case level' class of Qualifier 1 compared to `event level' suggests that utterances involving cases are more challenging. This is likely because case-level process mining concepts are often implicit in utterances, providing fewer contextual cues for the models; 
    \item the qualifier-level breakdown highlights challenges within the `domain' class of Qualifier 9 for both languages. Here, low performance is attributed to difficulties in linking utterance terms to the database schema, especially when domain-specific vocabulary is used; for example, terms like `declarations' in the utterance often lack a clear association with items in the schema. 
    \item although English generally performed slightly better than Portuguese on average, closer inspection of each qualifier class reveals situations where English faced greater challenges. For instance, in the `descriptive analysis' class of Qualifier 2, English shows a significant 26.7 drop in the structure indicator for paraphrased statements compared to Portuguese, underscoring the complexity of handling paraphrased utterances in this context. 
\end{itemize}

\begin{table}[htp]
  \caption{Results for the structure indicator considering the qualifiers from the process mining perspective. The results are presented as a percentage of correct answers along with the corresponding absolute numbers of each corresponding qualifier class, for Portuguese and English utterances.}
  \label{tab:results_pmp_structure}

\begin{tabular}{cp{2.7cm}ccccc}\toprule

\multirow{2}{*}{\textbf{Qualifier}} &\multirow{2}{*}{\textbf{Qualifier values}} &\multicolumn{2}{l}{Portuguese} & &\multicolumn{2}{l}{English} \\\cmidrule{3-4}\cmidrule{6-7}

& &\textbf{Base} &\textbf{Paraphrase} & &\textbf{Base} &\textbf{Paraphrase} \\\midrule

\multirow{3}{*}{1} &event level &$\underset{(41/95)}{43.2}$ &$\underset{(244/699)}{34.9}$ & &$\underset{(45/95)}{47.4}$ &$\underset{(252/699)}{36.1}$ \\
&case level &$\underset{(20/61)}{32.8}$ &$\underset{(105/435)}{24.1}$ & &$\underset{(17/61)}{27.9}$ &$\underset{(108/435)}{24.8}$ \\
& & & & & & \\

\multirow{3}{*}{2} &perspective &$\underset{(61/120)}{50.8}$ &$\underset{(345/821)}{42.0}$ & &$\underset{(59/120)}{49.2}$ &$\underset{(342/821)}{41.7}$ \\
&descriptive statistics &$\underset{(31/100)}{31.0}$ &$\underset{(171/721)}{23.7}$ & &$\underset{(33/100)}{33.0}$ &$\underset{(174/721)}{24.1}$ \\
&conformance &$\underset{(5/15)}{33.3}$ &$\underset{(18/109)}{16.5}$ & &$\underset{(6/15)}{40.0}$ &$\underset{(17/109)}{15.6}$ \\
& & & & & & \\

\multirow{3}{*}{9} &value &$\underset{(43/86)}{50.0}$ &$\underset{(226/493)}{45.8}$ & &$\underset{(43/86)}{50.0}$ &$\underset{(232/493)}{47.1}$ \\
&generic &$\underset{(17/66)}{25.8}$ &$\underset{(86/381)}{22.6}$ & &$\underset{(19/66)}{28.8}$ &$\underset{(98/381)}{25.7}$ \\
&domain &$\underset{(1/4)}{25.0}$ &$\underset{(37/260)}{14.2}$ & &$\underset{(0/4)}{0.0}$ &$\underset{(30/260)}{11.5}$ \\
\bottomrule
\end{tabular}
\end{table}

Tables \ref{tab:results_nlp_structure}  provides detailed results for the Qualifier 5 of NLP perspective. The structure indicator for Qualifier 5 highlights that the `who'  class presents a challenge in both languages. All utterances in this class also fall under the `domain' category of Qualifier 9, which, as previously discussed, suffers from domain-specific vocabulary challenges. The same utterances complexities aforementioned negatively impacted `what' and `which'  classes. Since most types of `what' and `which' statements are retrieving the complete tuple in projections, some generated SQL statements are considered unsuccessful because they explicitly mention each attribute in the PROJECT clause, while the gold SQL uses the alias `*'. 

\begin{table}[htp]
  \caption{Results for the structure indicator considering a qualifier from the natural language process perspective. The results are presented as a percentage of correct answers along with the corresponding absolute numbers of each corresponding qualifier class, for Portuguese and English utterances.}
  \label{tab:results_nlp_structure}
     \begin{tabular}{cp{1.4cm}ccccc}\toprule
     
        \multirow{2}{*}{\textbf{Qualifier}} &\textbf{Qualifier} &\multicolumn{2}{c}{Portuguese} & &\multicolumn{2}{c}{English} \\
        
        \cmidrule{3-4}\cmidrule{6-7}
        
        & \textbf{values} &\textbf{Base} &\textbf{Paraphrase} & &\textbf{Base} &\textbf{Paraphrase}\\
        
        \midrule
        
        \multirow{6}{*}{5} &how &$\underset{(21/53)}{39.6}$ &$\underset{(86/302)}{28.5}$ & &$\underset{(20/53)}{37.7}$ &$\underset{(78/302)}{25.8}$ \\
        
        &what &$\underset{(16/43)}{37.2}$ &$\underset{(55/197)}{27.9}$ & &$\underset{(17/43)}{39.5}$ &$\underset{(64/197)}{32.5}$ \\
        
        &none &$\underset{(16/31)}{51.6}$ &$\underset{(163/465)}{35.1}$ & &$\underset{(15/31)}{48.4}$ &$\underset{(169/465)}{36.3}$ \\
        
        &which &$\underset{(8/32}{25.0}$ &$\underset{(44/176)}{25.0}$ & &$\underset{(10/32)}{31.2}$ &$\underset{(48/176)}{27.3}$ \\
        
        &who &$\underset{(0/0)}{0.0}$ &$\underset{(0/7)}{0.0}$ & &$\underset{(0/0)}{0.0}$ &$\underset{(0/7)}{0.0}$ \\
        
        &when &$\underset{(0)}{0.0}$ &$\underset{(2/2)}{100.0}$ & &$\underset{(0/0)}{0.0}$ &$\underset{(2/2)}{100.0}$\\
        \bottomrule
\end{tabular}
\end{table}

Tables \ref{tab:results_sql_structure} provides detailed results for the Qualifier 8 from SQL perspective. This qualifier reveals a negative correlation between the complexity and structure indicator success. The `hard' and `extra-hard' classes have the lowest structure indicator values, as expected, due to their increased complexity.

\begin{table}[htp]
  \caption{Results for the structure indicator considering a qualifier from the structured query language perspective. The results are presented as a percentage of correct answers along with the corresponding absolute numbers of each corresponding qualifier class, for Portuguese and English utterances.} 
  \label{tab:results_sql_structure}
   \begin{tabular}{clcccccc}\toprule
        \multirow{2}{*}{\textbf{Qualifier}} &\textbf{Qualifier} &\multicolumn{2}{c}{Portuguese} & &\multicolumn{2}{c}{English} \\
        
        \cmidrule{3-4}\cmidrule{6-7}
        
        &\textbf{values} &\textbf{Base} &\textbf{Paraphrase} & &\textbf{Base} &\textbf{Paraphrase} \\
        
        \midrule
        
        \multirow{5}{*}{8} &easy &$\underset{(24/37)}{64.9}$ &$\underset{(143/285)}{50.2}$ & &$\underset{(27/37)}{73.0}$ &$\underset{(152/285)}{53.3}$ \\
        
        &medium &$\underset{(32/74)}{43.2}$ &$\underset{(167/546)}{30.6}$ & &$\underset{(29/74)}{39.2}$ &$\underset{(165/546)}{30.2}$ \\
        
        &hard &$\underset{(3/31)}{9.7}$ &$\underset{(28/220)}{12.7}$ & &$\underset{(4/31)}{12.9}$ &$\underset{(32/220)}{14.5}$ \\
        
        &extra &$\underset{(2/14)}{14.3}$ &$\underset{(11/83)}{13.3}$ & &$\underset{(2/14)}{14.3}$ &$\underset{(11/83)}{13.3}$ \\
        
        &no hardness &- &- & &- &- \\
        
        \bottomrule
        \end{tabular}
\end{table}

\vspace{0.2cm}
\noindent\textbf{\textit{Run indicator}}: Across all utterance-SQL pairs, the success rate for the run indicator was 44.5\% for Portuguese and 47.6\% for English, corresponding to 737 and 788 correct answers, respectively. The improvement in the run indicator relative to the structure indicator suggests that GPT-3.5 Turbo model generates SQL statements that may fail structural checks but still produce accurate results, as shown in the examples of Table \ref{tab:run_indicator_example}.

\begin{table}[htp]
  \caption{Examples of success on run indicator but fail on structure indicator}
  \label{tab:run_indicator_example}
  \begin{tabular}{p{0.4cm}p{5.5cm}p{5.5cm}}
    \toprule
    &Generated & Gold\\
    \midrule
    EN & SELECT AVG(cost) FROM event\_log WHERE activity = `payment handled' AND \textcolor{blue}{strftime (`\%Y', timestamp) $<$ `2018'} & SELECT AVG(cost) FROM event\_log WHERE \textcolor{blue}{timestamp $<$ `2018-01-01'} AND activity = `Payment handled'\\
    PT & SELECT activity, COUNT(*) FROM event\_log WHERE activity \textcolor{blue}{IN (`início da viagem', `fim da viagem')} GROUP BY activity; & SELECT count(*), activity from event\_log where \textcolor{blue}{activity = `Início da viagem' OR activity = `Fim da viagem'} GROUP BY activity\\
    \bottomrule
\end{tabular}
\end{table}

The results for the run indicator detailed per qualifier are depicted in tables \ref{tab:results_pmp_run}, \ref{tab:results_nlp_run}, and \ref{tab:results_sql_run}. With regard to the analysis of the PMp qualifiers, it is observed that there was a significant improvement in the value of the run indicator, in relation to the structure indicator, for utterances at the `event level', for `descriptive statistics', and for general-purpose ones (`generic'). This may indicate both an ease in the correct interpretation of the utterances by the GPT-3.5 Turbo model and the possibility of task resolution with a broader variety of equivalent statements (although syntactically different). It is also observed that higher values in the run indicator are obtained for utterances that address the value `how' for Qualifier 5, and for all levels of SQL statement complexity (Qualifier 8). Furthermore, in this indicator, it is possible to evaluate the SQL classifiers as `no hardness', for which the GPT-3.5 Turbo model shows the greatest difficulty in resolving the text-to-SQL problem. 

\begin{table}[htp]
  \caption{Results for the run indicator considering the qualifier organization of utterances from the process mining perspective. The results are presented as a percentage of correct answers along with the corresponding absolute numbers of each corresponding qualifier class, for Portuguese and English utterances.}
  \label{tab:results_pmp_run}

    \begin{tabular}{cp{2.7cm}ccccc}\toprule
    
        \multirow{2}{*}{\textbf{Qualifier}} &\multirow{2}{*}{\textbf{Qualifier values}} &\multicolumn{2}{c}{Portuguese} & &\multicolumn{2}{c}{English} \\
        
        \cmidrule{3-4}\cmidrule{6-7}
        
        & &\textbf{Base} &\textbf{Paraphrase} & &\textbf{Base} &\textbf{Paraphrase} \\
        
        \midrule
        
        \multirow{2}{*}{1} &event level &$\underset{(71/110)}{64.5} $&$\underset{(427/802)}{53.2}$ & &$\underset{(75/110)}{68.2}$ &$\underset{(451/802)}{56.2}$ \\
        &case level &$\underset{(37/95)}{38.9}$ &$\underset{(202/648)}{31.2}$ & &$\underset{(38/95)}{40.0}$ &$\underset{(224/648)}{34.6}$ \\
        & & & & & & \\
        
        \multirow{3}{*}{2} &perspective &$\underset{(78/165)}{47.3}$ &$\underset{(388/1130)}{34.3}$ & &$\underset{(83/165)}{50.3}$ &$\underset{(432/1130)}{38.2}$ \\
        &descriptive statistics &$\underset{(79/129)}{61.2}$ &$\underset{(466/894)}{52.1}$ & &$\underset{(80/129)}{62.0}$ &$\underset{(478/894)}{53.5}$ \\
        &conformance &$\underset{(9/24)}{37.5}$ &$\underset{(42/180)}{23.3}$ & &$\underset{(9/24)}{37.5}$ &$\underset{(39/180)}{21.7}$ \\
        & & & & & & \\
        
        \multirow{3}{*}{9} &value &$\underset{(58/115)}{50.4}$ &$\underset{(323/623)}{51.8}$ & &$\underset{(62/115)}{53.9}$ &$\underset{(355/623)}{57.0}$ \\
        &generic &$\underset{(49/84)}{58.3}$ &$\underset{(272/484)}{56.2}$ & &$\underset{(51/84)}{60.7}$ &$\underset{(288/484)}{59.5}$ \\
        &domain &$\underset{(1/6)}{16.7}$ &$\underset{(34/343)}{9.9}$ & &$\underset{(0/6)}{0.0}$ &$\underset{(32/343)}{9.3}$ \\
        \bottomrule
    \end{tabular}
\end{table}

\begin{table}[htp]
  \caption{Results for the run indicator considering the qualifier organization of utterances from the natural language process perspective. The results are presented as a percentage of correct answers along with the corresponding absolute numbers of each corresponding qualifier class, for Portuguese and English utterances.}
  \label{tab:results_nlp_run}
     \begin{tabular}{cp{1.4cm}ccccc}\toprule
     
        \multirow{2}{*}{\textbf{Qualifier}} &\textbf{Qualifier} &\multicolumn{2}{c}{Portuguese} & &\multicolumn{2}{c}{English} \\
        
        \cmidrule{3-4}\cmidrule{6-7}
        
        & \textbf{values} &\textbf{Base} &\textbf{Paraphrase} & &\textbf{Base} &\textbf{Paraphrase}\\
        
        \midrule
        
        \multirow{6}{*}{5} &how &$\underset{(49/72)}{68.1}$ 
        &$\underset{(216/401)}{53.9}$ & &$\underset{(49/72)}{68.1}$ 
         &$\underset{(219/401)}{54.6}$ \\
        
        &what &$\underset{(24/56)}{42.9}$ &$\underset{(88/257)}{34.2}$ &
        &$\underset{(27/56)}{48.2}$ &$\underset{(105/257)}{40.9}$ \\
        
        &none &$\underset{(23/40)}{57.5}$
        &$\underset{(276/583)}{47.3}$ &
        &$\underset{(23/40)}{57.5}$
        &$\underset{(295/583)}{50.6}$\\
        
        &which &$\underset{(12/40)}{30.0}$ &$\underset{(55/217)}{25.3}$ & &$\underset{(15/40)}{37.5}$ &$\underset{(64/217)}{29.5}$ \\
        
        &who &$\underset{(0/0)}{0.0}$ &$\underset{(0/8)}{0.0}$ & &$\underset{(0/0)}{0.0}$ &$\underset{(0/8)}{0.0}$ \\
        
        &when &$\underset{(0)/0}{0.0}$ &$\underset{(2/2)}{100.0}$ & &$\underset{(0/0)}{0.0}$ & $\underset{(2/2)}{100.0}$ \\
        
        \bottomrule
    \end{tabular}
\end{table}

\begin{table}[htp]
  \caption{Results for the run indicator considering the qualifier organization of utterances from the structured query language perspective. The results are presented as a percentage of correct answers along with the corresponding absolute numbers of each corresponding qualifier class, for Portuguese and English utterances.} 
  \label{tab:results_sql_run}
   \begin{tabular}{clcccccc}\toprule
        \multirow{2}{*}{\textbf{Qualifier}} &\textbf{Qualifier} &\multicolumn{2}{c}{Portuguese} & &\multicolumn{2}{c}{English} \\
        
        \cmidrule{3-4}\cmidrule{6-7}
        
        &\textbf{values} &\textbf{Base} &\textbf{Paraphrase} & &\textbf{Base} &\textbf{Paraphrase} \\
        
        \midrule
        
        \multirow{5}{*}{8} &easy &$\underset{(27/37)}{73.0}$ &$\underset{(156/285)}{54.7}$ & &$\underset{(32/37)}{86.5}$ &$\underset{(183/285)}{64.2}$ \\
        
        &medium &$\underset{(54/74)}{73.0}$ &$\underset{(345/546)}{63.2}$ & &$\underset{(55/74)}{74.3}$ &$\underset{(359/546)}{65.8}$ \\
        
        &hard &$\underset{(11/31)}{35.5}$ &$\underset{(66/220)}{30.0}$ & &$\underset{(10/31)}{32.3}$ &$\underset{(74/220)}{33.6}$ \\
        
        &extra &$\underset{(5/14)}{35.7}$ &$\underset{(25/83)}{30.1}$ & &$\underset{(5/14)}{35.7}$ &$\underset{(23/83)}{27.7}$ \\
        
        &no hardness &$\underset{(11/49)}{22.4}$ &$\underset{(37/316)}{11.7}$ & &$\underset{(11/49)}{22.4}$ &$\underset{(36/316)}{11.4}$ \\
        
        \bottomrule
        \end{tabular}
\end{table}

\vspace{0.2cm}
\noindent\textbf{\textit{Comparative analysis}}: An analysis based on the GPT-3.5 Turbo model's performance from the perspective of Qualifier 8 (SQL perspective) confirms that although more complex SQL statements are harder for the GPT-3.5 Turbo model to generate, there are also more ways to design favoring the evaluation under the run indicator (there is a significant difference between the values obtained in the structure indicator and the run indicator for queries at the hard and extra complexity levels).

\begin{figure}[htp]
  \centering
  \includegraphics[width=\linewidth]{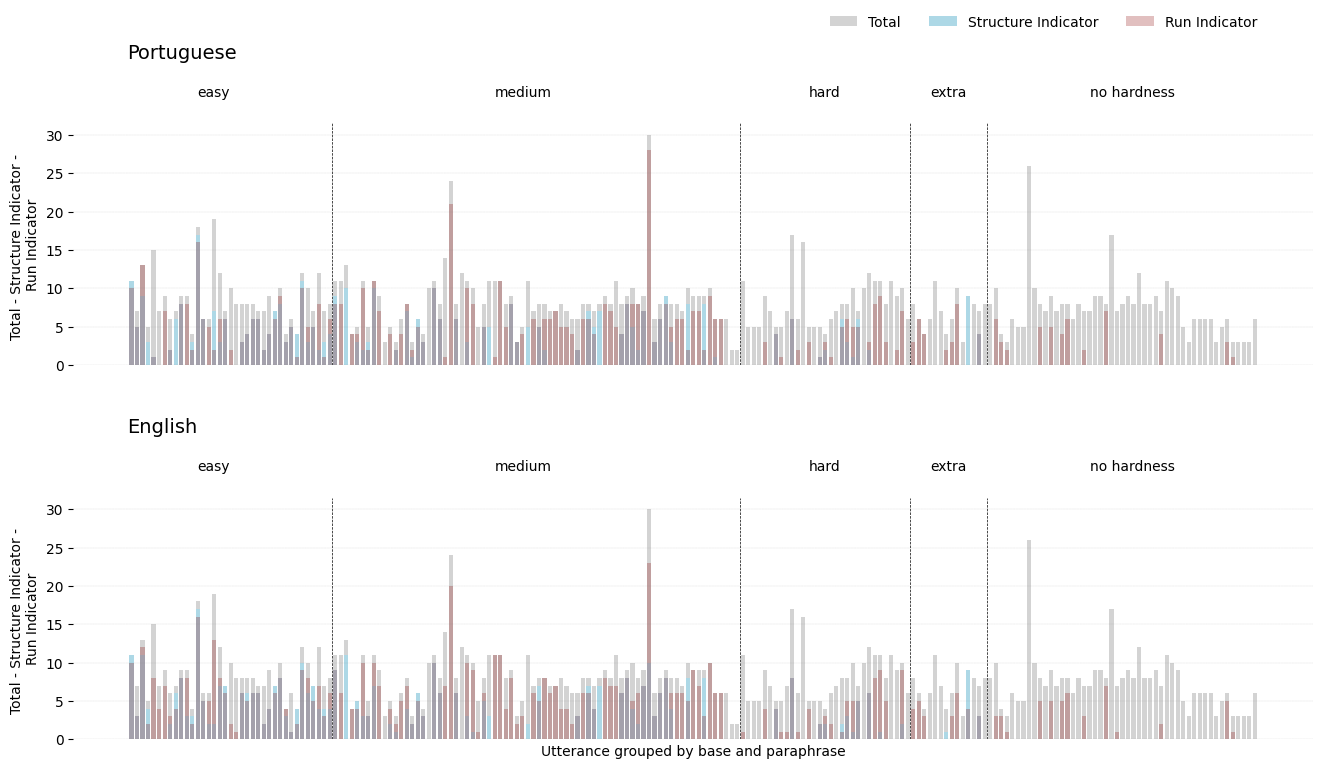}
  \caption{Distribution of success rates for translating natural language utterances into SQL statements, grouped by base utterance and corresponding paraphrases. Each bar in the chart represents a group. The bar size indicates the number of utterances in each group (base utterance + corresponding paraphrases). Bars entirely in light gray indicate that the translation failed for all utterances in the group. Dark gray in a bar indicates the number of utterances for which the translation succeeded in both indicators. Blue and red colors indicate that the translation succeeded in only one indicator, structure indicator or run indicator, respectively. The qualifier $8$, related to the complexity of the expected SQL statement, is used to organize the groups under analysis.}
  \label{fig:hardness_success_by_utterance}
\end{figure}

Grouping each base utterance with their paraphrases and segregating by hardness classes, the distribution on figure \ref{fig:hardness_success_by_utterance} shown that the GPT-3.5 Turbo model fails according to both indicator for only 27 and 23 utterance-SQL pairs (sections shaded in light gray on the bars in the figure), for Portuguese and English respectively. Specifically for the unclassified pairs (`no hardness'), evaluated only under the run indicator, failures are observed more frequently (37 and 36 for Portuguese and English, respectively).

\vspace{0.2cm}
\noindent\textbf{\textit{General challenges}}: The process mining domain has specific characteristics that pose challenges for text-to-SQL solutions. In the presented \NOME baseline, three characteristics stand out, and they are illustrated in tables \ref{tab:examples_utterance_complexities_process_mining_A}, \ref{tab:examples_utterance_complexities_process_mining_B}, and \ref{tab:examples_utterance_complexities_process_mining_C}:

\begin{itemize}
    \item in the example in Table \ref{tab:examples_utterance_complexities_process_mining_A}, the understanding regarding the number of instances that can be associated with the object to be retrieved (suggested by the plural term `cases' -- more than one case can have the `largest' size in terms of the number of events) is correctly expressed in the gold SQL statement but not addressed in the generated SQL statement (which necessarily returns a single instance as the result). Although this problem can be found in other application domains, it becomes especially relevant in process mining when retrieving information at the `case level', the main object of interest in process mining. 
    \item in the example in Table \ref{tab:examples_utterance_complexities_process_mining_B}, the understanding of the temporal order of events (`permit final approved by supervisor' must occur before `start trip'), correctly expressed in the gold SQL statement, but incorrectly represented in both generated SQL statements. In this case, for English, no order was required in the result, and for Portuguese, the reverse order was required. Information related to the sequence of events is especially important in process mining due to the interest in understanding the execution dynamics of a business process.
    \item in the examples of Table \ref{tab:examples_utterance_complexities_process_mining_C}, the GPT-3.5 model provides SQL statements that partially address the expected answer (first example on table); or does not deal with the DISTINCT command (second example on table). The latter problem stems from the fact that the data in the event log is not normalized (`case' attribute) or not unique (`activity' attribute). In both cases, one could say that there is a problem of dubiety in natural language, as the result obtained from executing the generated SQL statements can be considered correct under some level of abstraction in the interpretation of the utterance.    
    \end{itemize}

\begin{table}[htp]
  \caption{Example of utterance with difficult for the text-to-SQL solution: number of instances to be recovered. The main differences between the gold and generated SQL statements are highlighted in blue.}
  \label{tab:examples_utterance_complexities_process_mining_A}
  \begin{tabular}{p{0.4cm}p{12cm}}
    \toprule
    EN & \textbf{Utterance}: Retrieve the identifiers of the cases that have the greatest number of events.\\
    PT & \textbf{Utterance}: \textit{Recupere os identificadores dos casos que possuem a maior quantidade de eventos.}\\
       & \textbf{Gold SQL}: \\ 
       & SELECT idcase FROM event\_log GROUP BY idcase HAVING COUNT(*) = (SELECT COUNT(*) FROM event\_log GROUP BY idcase ORDER BY count(*) DESC LIMIT 1)\\
        & \textbf{Generated SQL - for both languages}: \\
        &  SELECT idcase FROM event\_log GROUP BY idcase \textcolor{blue}{ORDER BY COUNT(id) DESC LIMIT 1} \\
    \bottomrule
\end{tabular}
\end{table}

\begin{table}[htp]
  \caption{Example of utterance with difficult for the text-to-SQL solution: temporal ordering. The main differences between the gold and generated SQL statements are highlighted in blue.}
  \label{tab:examples_utterance_complexities_process_mining_B}
  \begin{tabular}{p{0.4cm}p{12cm}}
    \toprule
     EN & \textbf{Utterance}: In how many cases does the `start trip' activity eventually follow the `permit final approved by supervisor' activity?\\
      PT & \textbf{Utterance}: \textit{Em quantos casos a atividade `início da viagem' segue a atividade `permissão final aprovada pelo supervisor' eventualmente?} \\
        & \textbf{Gold SQL}: SELECT COUNT(DISTINCT e1.idcase) FROM event\_log e1 INNER JOIN event\_log e2 ON e1.idcase = e2.idcase WHERE e1.activity = `Permit final approved by supervisor' AND e2.activity = `Start trip' AND e1.timestamp $<$ e2.timestamp\\
       & \textbf{Generated SQL - for English}: \\
         & SELECT COUNT(DISTINCT idcase) FROM event\_log \textcolor{blue}{WHERE activity = `start trip' AND idcase IN ( SELECT idcase FROM event\_log WHERE activity = `permit final approved by supervisor')} \\
         & \textbf{Generated SQL - for Portuguese}: \\ 
       & SELECT COUNT(DISTINCT \textcolor{blue}{idcase}) FROM event\_log e1 \textcolor{blue}{JOIN} event\_log e2 ON e1.idcase = e2.idcase WHERE e1.activity = \textcolor{blue}{`início da viagem'} AND e2.activity = \textcolor{blue}{`permissão final aprovada pelo supervisor'} AND e1.timestamp $<$ e2.timestamp\\
    \bottomrule
\end{tabular}
\end{table}

\begin{table}[htp]
  \caption{Example of utterance with difficult for the text-to-SQL solution: incompleteness or dubiety. The main differences between the gold and generated SQL statements are highlighted in blue.}
  \label{tab:examples_utterance_complexities_process_mining_C}
  \begin{tabular}{p{0.4cm}p{12cm}}
    \toprule
    EN & \textbf{Utterance}: What are the events that are not associated with resources?\\
    PT & \textbf{Utterance}: \textit{Quais são os eventos que não estão associados a recursos?}\\
       & \textbf{Gold SQL}: SELECT * FROM event\_log WHERE resource IS NULL\\
       & \textbf{Generated SQL - for both languages}: \\
       & SELECT \textcolor{blue}{activity, timestamp, cost, idcase} FROM event\_log WHERE resource IS NULL \\
    & \\
    EN & \textbf{Utterance}: Which activities contain the word `declaration' in their label?\\
    PT & \textbf{Utterance}: \textit{Quais atividades contém a palavra `declaração' em seu rótulo?}\\
        & \textbf{Gold SQL}: SELECT DISTINCT activity FROM event\_log WHERE activity LIKE `\%declaration\%'\\
        & \textbf{Generated SQL - for both languagems}: \\
        & SELECT \textcolor{blue}{activity} FROM event\_log WHERE activity LIKE `\%declaration\%'; \\
    \bottomrule
\end{tabular}
\end{table}

\section{Conclusion}
\label{sec:conclusions}

In this paper, we present the benchmark dataset \NOME. \NOME has been established as a benchmark because it offers both the dataset and a baseline solution for the text-to-SQL task. This dataset is specifically tailored to the process mining domain -- a prominent area of data exploration where MANY stakeholders lack technical expertise in SQL but seek to retrieve information related to business process management.

The primarily use case for \textit{PM$_{text2sql}$} dataset was for fine-tuning and evaluation of text-to-SQL implementations in process mining context for request information; therefore, it can be used for other natural language tasks because their richness features such as: i) bilingual, ii) a curated and assessment process of the generated utterances and corresponding SQL statements, iii) a careful enrichment with paraphrases and iv) qualifiers from diverse perspectives; and (v) careful curation by humans. The bilingual nature of the dataset, featuring natural language utterances and enriched paraphrases generated by native Portuguese speakers, along with their corresponding English versions translated by a professional, makes it a valuable resource for semantic parsing tasks and potentially for other natural language processing tasks such as machine translation or paraphrase generation. Specifically, within the process mining domain, the carefully crafted paraphrases created by researchers with extensive expertise in this area represent an especially valuable resource.

The particular features of the \NOME dataset for the text-to-SQL task are: 

\begin{itemize}
\item \NOME contains Portuguese natural language utterances and the corresponding values in SQL statements also in Portuguese, so that can be used to fine-tune and evaluate text-to-SQL implementations in Portuguese language. \NOME joins the mRAT-SQL+GAP dataset created using automatic translations of the Spider dataset from English to Portuguese \cite{archanjo_2021_mRATSQL_GAP}, serving as another resource for automatic processing of Portuguese.

\item although the simple one-table schema structure of a event log has been used, the request information that needs timestamped sequence of events to be answered, so much common in process mining context, can be a quite challenge to construct a SQL statement. Thus, some of the natural language utterances of the dataset imposes a real challenge to text-to-SQL implementations and serves as a precise benchmark in such a domain.

\item the utterances manual qualification can be used for classification tasks or a separation of concerns when assessing a model implementation with the \NOME dataset. 

\end{itemize}

The limitations of the \NOME benchmark dataset are:

\begin{itemize}
\item it can only be used to train text-to-SQL models for exploratory information retrieval tasks. Other types of information requests, which are important in process mining but require advanced event log processing - such as discovery, optimization, and process monitoring — are not covered. This limitation can only be partially overcome with future research efforts due to inherent constraints of the SQL language itself. An alternative to extend the utility of semantic parsing in process mining, though not fully overcoming the limitation at hand, could be to focus on creating datasets using a process query language \cite{vogelgesang_2022_celonis_pql}.
\item the SQLite DBMS was used for verification and validation of the statements SQL. Therefore, some pairs of utterance-SQL statement in the dataset use specific SQLite functions, mainly date time and window functions.  
\end{itemize}

\bibliography{sn-bibliography} 

\end{document}